\documentclass[aps, preprint, prd, showpacs, showkeys, nofootinbib, superscriptaddress]{revtex4}
\usepackage{times}
\usepackage{amssymb,amsbsy,amsmath,amsfonts}
\usepackage{graphicx}
\usepackage{float}
\usepackage{color}
\usepackage{morefloats}
\usepackage{rotating}
\usepackage{srcltx}
\usepackage{slashed}
\usepackage{multirow}
\usepackage{verbatim}
\usepackage{tabularx}
\renewcommand{\arraystretch}{1.2}
\newcommand{\PreserveBackslash}[1]{\let\temp=\\#1\let\\=\temp}
\newcolumntype{C}[1]{>{\PreserveBackslash\centering}p{#1}}
\newcolumntype{R}[1]{>{\PreserveBackslash\raggedleft}p{#1}}
\newcolumntype{L}[1]{>{\PreserveBackslash\raggedright}p{#1}}

\begin{document}

\title{Strangeness $S=-1$ hyperon-nucleon scattering in covariant chiral effective field theory}

\author{Kai-Wen Li}
\affiliation{School of Physics and Nuclear Energy Engineering and International Research Center for Nuclei and Particles
in the Cosmos, Beihang University, Beijing 100191, China}

\author{Xiu-Lei Ren}
\affiliation{School of Physics and Nuclear Energy Engineering and International Research Center for Nuclei and Particles
in the Cosmos, Beihang University, Beijing 100191, China}
\affiliation{School of Physics and State Key Laboratory of Nuclear Physics and Technology, Peking University, Beijing 100871, China}

\author{Li-Sheng Geng}
\email[E-mail me at: ]{lisheng.geng@buaa.edu.cn}
\affiliation{School of Physics and Nuclear Energy Engineering and International Research Center for Nuclei and Particles
in the Cosmos, Beihang University, Beijing 100191, China}
\affiliation{Beijing Key Laboratory of Advanced Nuclear Materials and Physics, Beihang University, Beijing 100191, China}

\author{Bingwei Long}
\affiliation{Center for Theoretical Physics, Department of Physics, Sichuan University, 29 Wang-Jiang Road, Chengdu, Sichuan 610064, China}

\begin{abstract}
Motivated by the successes of covariant baryon chiral perturbation theory in one-baryon systems and in heavy-light systems, we study relevance of relativistic effects in hyperon-nucleon interactions with strangeness $S=-1$. In this exploratory work,
we follow the covariant framework developed by Epelbaum and Gegelia to calculate the $YN$ scattering amplitude at leading order. By fitting the five low-energy constants to
the experimental data, we find that  the cutoff dependence is mitigated, compared with the heavy-baryon approach. Nevertheless, the description of the experimental data remains quantitatively similar at
leading order.

\end{abstract}


\date{\today}

\maketitle
\section{Introduction}

Hyperon-nucleon ($YN$) interactions play an important role in our understanding of
hypernuclear physics and neutron stars~\cite{Nogga:2001ef,Lonardoni:2014bwa}.  Because of the nonperturbative nature of strong interactions,
previous theoretical investigations were based on either meson-exchange models or quark models, such as the Nijmegen meson-exchange model~\cite{ Nagels:1976xq,Maessen:1989sx,Rijken:1998yy,Rijken:2006ep,Nagels:2015lfa},
 the J\"ulich meson-exchange model~\cite{Holzenkamp:1989tq,Haidenbauer:2005zh},  the T\"ubingen quark cluster model
\cite{Straub:1988gj, Straub:1990de,Zhang:1994pp} , the chiral SU(3) quark model~\cite{Zhang:1997ny},   the quark delocalization and color screening model~\cite{Ping:1998si}, and the Kyoto-Niigata SU(6) quark cluster model~\cite{Fujiwara:1995fx,Fujiwara:2006yh}. We will study here $YN$ scattering using a covariant-baryon framework of chiral effective field theory (ChEFT).

Weinberg first proposed in the 1990s using techniques of ChEFT to develop nucleon-nucleon ($NN$) interactions~\cite{Weinberg:1990rz, Weinberg:1991um}, and impressive progress has been made along this line of research~\cite{Bedaque:2002mn,Epelbaum:2008ga,Machleidt:2011zz}. Generalization of the method included antinucleon-nucleon~\cite{Kang:2013uia}, hyperon-hyperon~\cite{Polinder:2007mp,Haidenbauer:2009qn,Haidenbauer:2014rna,Haidenbauer:2015zqb}, and hyperon-nucleon interactions~\cite{Polinder:2006zh,Haidenbauer:2007ra,Haidenbauer:2013oca,Petschauer:2013uua}---our focus in the present paper. The main advantage of this approach is that the description of experimental data can be systematically improved by calculating higher orders following a power counting scheme. In addition, three- and higher-body forces can be treated in the same framework as two-body forces. Furthermore, theoretical uncertainties can be systematically estimated if the power counting is consistent.

In Weinberg's original proposal, and in accordance with the conventional practice in low-energy nuclear physics, baryons are treated as nonrelativistic objects at leading order (LO), with relativistic corrections accounted for in higher corders. The machinery to implement this idea is the heavy-baryon (HB) formalism~\cite{Jenkins_1990jv}.

Another important ingredient is the assumption of four-baryon coupling constants conforming to naive dimensional analysis (NDA) so that derivatives and quark-mass dependence in vertexes are always suppressed by $\Lambda_\chi \sim 1$ GeV, the breakdown scale of ChEFT. We will refer to this scheme, HB plus NDA, as the HB approach throughout the paper. The premise of NDA was challenged by some authors, for NDA does not assign sufficient baryon-baryon contact potentials to remove ultraviolet cutoff dependence from the $NN$ scattering amplitudes, even at LO. Stated differently, one does not seem to be able to reconcile NDA with renormalization-group invariance.

Partly as an attempt to redeem NDA, Epelbaum and Gegelia proposed in their recent papers~\cite{Epelbaum:2012ua,Epelbaum:2015sha} a covariant ChEFT framework for $NN$ scattering, referred to as the EG approach in the present paper. While retaining NDA, this approach uses a particular three-dimensional reduction of the relativistic Bethe-Salpeter equation to account for the propagation of two-nucleon intermediate states. One must pay necessary attention to subleading orders so as not to double-count or to miss relativistic effects. The cutoff sensitivity in all partial waves was removed except $^3 P_0$, where a nominally higher-order contact term was introduced to achieve renormalization-group invariance~\cite{Epelbaum:2012ua}. For other works on renormalization of chiral nuclear forces, see Refs.~\cite{Lepage:1997cs,PavonValderrama:2005gu,Birse:2005um, Nogga:2005hy,Epelbaum:2006pt,Long:2007vp,Epelbaum:2009sd,Yang:2009pn,Valderrama:2009ei,Valderrama:2011mv,Long:2011xw}.

Besides the possibility to ameliorate cutoff sensitivities, covariant treatment of baryons is intriguing for it describes data more efficiently, in the sense that it entails fewer
terms at higher orders than its HB counterparts, e.g., in the one-baryon sector~\cite{Geng:2008mf,Geng:2009ik,Geng:2011wq,Ren:2012aj,Ren:2014vea}, and its generalization to describe heavy hadrons in heavy-light systems with three light flavors \cite{Geng:2010vw,Geng:2010df,Altenbuchinger:2011qn} (see Ref.~\cite{Geng:2013xn} for a short review).  Indeed, the EG approach was shown to improve the description of the $NN$ scattering phase shifts, up to the orders considered~\cite{Epelbaum:2012ua, Epelbaum:2015sha}, in comparison with the HB approach (including the Kaplan-Savage-Wise scheme~\cite{Kaplan:1998tg}). These phenomenological successes are particularly encouraging in studying $YN$ scattering, because one cannot afford to employ as many undetermined low-energy constants as in $NN$, due to the fact that the data on $YN$ scattering are scarce and in most cases of relatively low quality. (On a side note, the analysis in Ref.~\cite{Lv_2016slh} showed that there are indeed theoretical rationales for covariant baryons in the one-baryon sector, at least in certain kinematic regions.)

In the present paper, we apply the EG approach to $YN$ scattering, that is, NDA for $YN$ contact terms and covariant formulation for baryon propagations, with the focus on $\Lambda N - \Sigma N$. The purpose of this exploratory work is twofold: to investigate (\textit{a}) whether the EG approach reduces cutoff dependence as it did for $NN$ and (\textit{b}) how much it improves the fit to the $YN$ scattering data.

A second important approach to study baryon-baryon interactions must be mentioned before we move on to further discussion on ChEFT-based $YN$ interactions. Lattice QCD simulations~\cite{Beane:2006gf,Nemura:2008sp,Beane:2009py,Inoue:2010hs,Inoue:2010es,Beane:2010hg,Beane:2011iw,Buchoff:2012ja,Beane:2012ey,Beane:2012vq,  Etminan:2014tya,Yamada:2015cra,Sasaki:2015ifa} provide an \textit{ab initio} numerical solution to QCD with quark and gluon degrees of freedom. Thanks to increasingly available computing resources and ever-evolving numerical algorithms, lattice QCD simulations have begun to play an indispensable role in determining baryon-baryon forces~\cite{Doi:2015oha,Doi:2015uvd}.

The paper is organized as follows. In Sec.~\ref{sec_formalism}, we briefly explain the formalism, including the derivation of the kernel potentials and the Kadyshevsky equation, which will be used to iterate the potentials. In Sec. III, the fitting procedure is explained in detail. Results and discussions are presented in Sec. IV, followed by a short summary and outlook in Sec. V.

\begin{figure}
  \centering
  \includegraphics[width=0.15\textwidth]{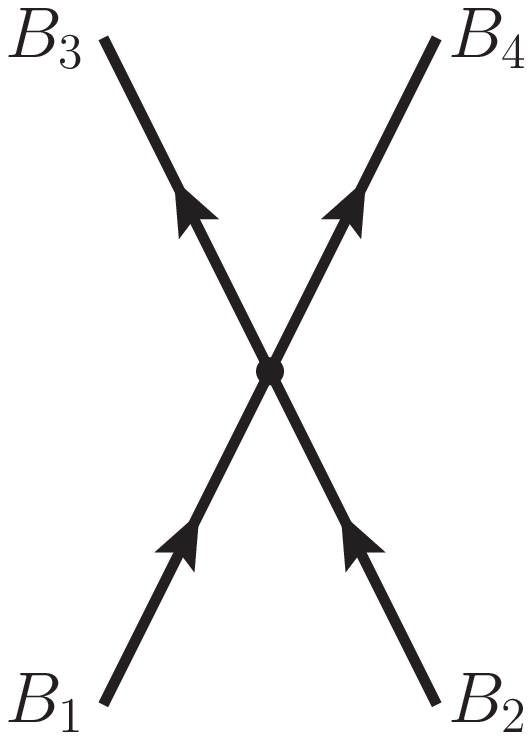}~~~~~~\qquad\qquad
  \includegraphics[width=0.15\textwidth]{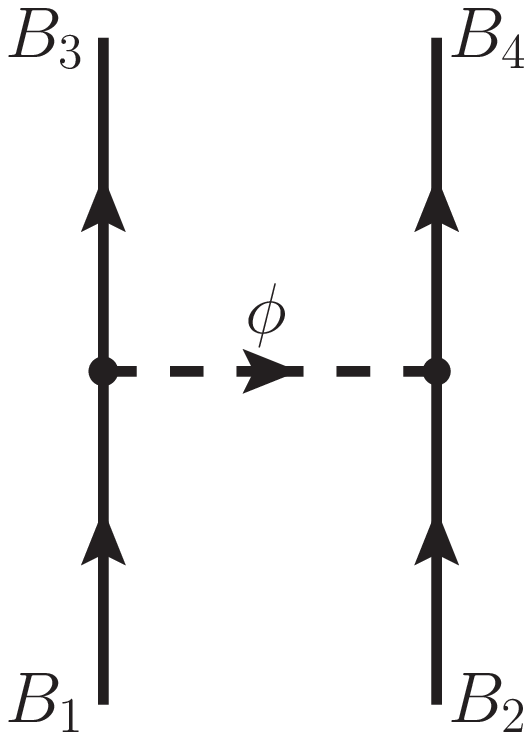}
  \caption{Nonderivative four-baryon contact terms and one-pseudoscalar-meson exchanges at LO. The solid lines denote incoming and outgoing baryons ($B_{1,2,3,4}$), and the dashed line
  denotes the exchanged pseudoscalar meson $\phi$.}\label{CTOME}
\end{figure}

\section{Formalism\label{sec_formalism}}

The LO potentials include nonderivative four-baryon contact terms and one-pseudoscalar-meson exchanges (OPME), as shown in Fig.~\ref{CTOME}. In both EG and HB approaches, LO potentials are obtained by applying on-shell conditions to external baryon lines in these diagrams; the difference between two approaches is the integral equation to iterate the potentials, as we will see soon. As far as the notations are concerned, we follow closely Ref.~\cite{Polinder:2006zh}.

\subsection{LO potentials}

The four-baryon contact terms have the form
\begin{align}\label{CT}
  &\mathcal{L}_{\textrm{CT}}^1 = C_i^1~\textrm{tr}\left(\bar B_a \bar B_b (\Gamma_i B)_b (\Gamma_i B)_a\right)\, ,
  \qquad
  \mathcal{L}_{\textrm{CT}}^2 = C_i^2~\textrm{tr}\left(\bar B_a (\Gamma_i B)_a \bar B_b (\Gamma_i B)_b\right)\, , \nonumber\\
  &
  \mathcal{L}_{\textrm{CT}}^3 = C_i^3~\textrm{tr}\left(\bar B_a (\Gamma_i B)_a\right)\textrm{tr}\left( \bar B_b (\Gamma_i B)_b\right)\, ,
\end{align}
where $\mathrm{tr}$ indicates trace in flavor space ($u$, $d$, and $s$); $\Gamma_i$ are the elements of the Clifford algebra,
\begin{equation}\label{CA}
  \Gamma_1=1\, ,~~~~~~\Gamma_2=\gamma^\mu\, ,~~~~~~\Gamma_3=\sigma^{\mu\nu}\, ,
  ~~~~~~\Gamma_4=\gamma^{\mu}\gamma_5\, ,~~~~~~\Gamma_5=\gamma_5\, ;
\end{equation}
and $C_i^m$ ($m=1,2,3$) are the low-energy constants (LECs) corresponding to independent four-baryon operators. The  ground-state octet baryons are collected in the $3\times3$ traceless matrix:
\begin{equation}\label{BMatrix}
  B =
  \left(
   \begin{array}{ccc}
    \frac{\Sigma^0}{\sqrt{2}}+\frac{\Lambda}{\sqrt{6}} & \Sigma^+ & p \\
    \Sigma^- & -\frac{\Sigma^0}{\sqrt{2}}+\frac{\Lambda}{\sqrt{6}} & n\\
    \Xi^- & \Xi^0 & -\frac{2\Lambda}{\sqrt{6}}
  \end{array}
  \right)\, .
\end{equation}

\begin{figure}
  \centering
  \includegraphics[width=0.15\textwidth]{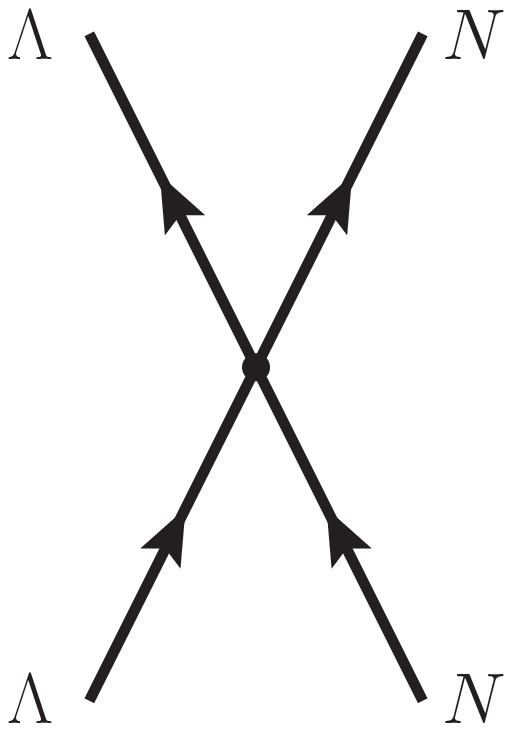}~~
  \includegraphics[width=0.15\textwidth]{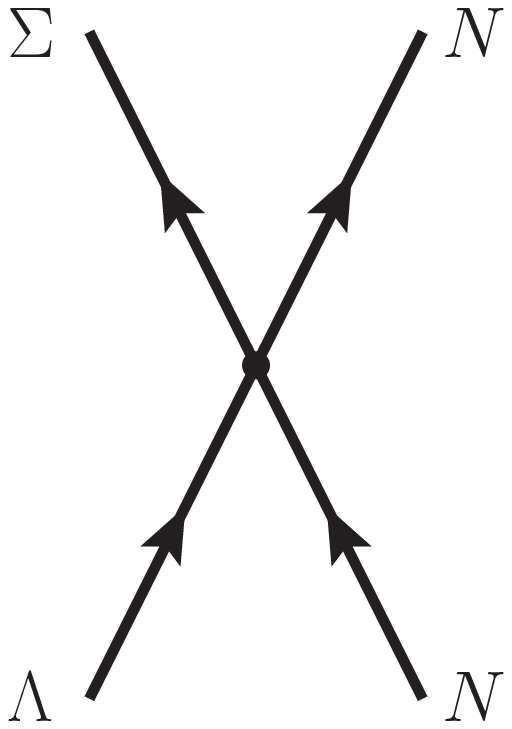}~~
  \includegraphics[width=0.15\textwidth]{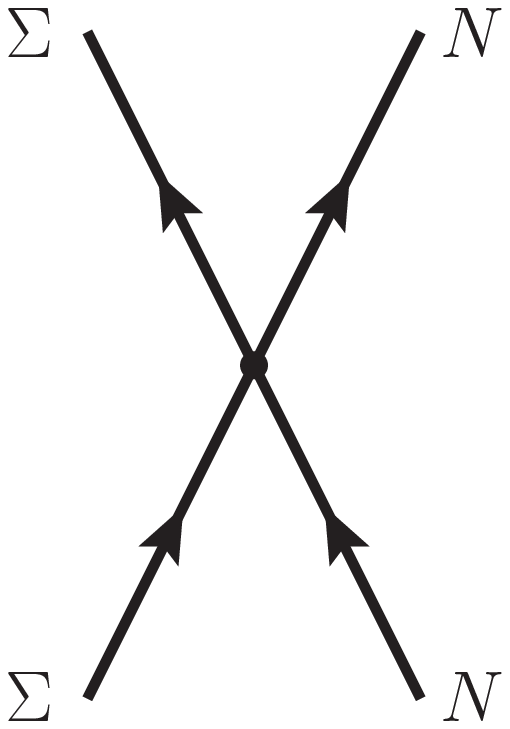}
  \caption{Nonderivative four baryon contact diagrams for the $\Lambda N-\Sigma N$ system.}\label{CT3}
\end{figure}

% \par
The OPME potentials are derived from the covariant $SU(3)$ meson-baryon Lagrangian,
\begin{equation}\label{LMB1}
  \mathcal{L}_{MB}^{(1)} =
  \mathrm{tr}\left( \bar B \big(i\gamma_\mu D^\mu - M_B \big)B  -\frac{D}{2} \bar B \gamma^\mu\gamma_5\{u_\mu,B\}
  -  \frac{F}{2}\bar{B} \gamma_\mu\gamma_5 [u_\mu,B]\right)\, ,
\end{equation}
where $D^\mu B = \partial_\mu B+[\Gamma_\mu,B] $, $M_B$ stands for the chiral limit mass of the octet baryons, and $D$ and $F$ are the axial vector couplings. In the numerical analysis, we will use
%$D+F(\equiv g_A)=1.26$
$D+F = g_A =1.26$ and $F/(F+D)=0.4$, where $g_A$ is the nucleon axial vector coupling constant. $\Gamma_\mu$ ($u_\mu$) are the vector and (axial vector) combinations of the pseudoscalar-meson fields and their derivatives,
\[
  \Gamma_\mu = \frac{1}{2}\left[u^\dag\partial_\mu u + u\partial_\mu u^\dag \right], \quad u_\mu=i(u^\dagger \partial_\mu u-u\partial_\mu u^\dagger)\, ,
\]
where $u^2= U = \exp\left(i\frac{\sqrt{2}\phi}{f_0}\right)$, with the pseudoscalar-meson decay constant $f_0=93$ MeV~\cite{Haidenbauer:2013oca}, and the traceless matrix $\phi$ collecting the pseudoscalar-meson fields:
\begin{equation}\label{MMatrix}
  \phi =
  \left(
   \begin{array}{ccc}
    \frac{\pi^0}{\sqrt{2}}+\frac{\eta}{\sqrt{6}} & \pi^+ & K^+ \\
    \pi^- & -\frac{\pi^0}{\sqrt{2}}+\frac{\eta}{\sqrt{6}} & K^0\\
    K^- & \bar K^0 & -\frac{2\eta}{\sqrt{6}}
  \end{array}
  \right).
\end{equation}

The LO potentials can be written schematically as
\begin{equation}
V_\mathrm{LO}=C^S_{B_1B_2\rightarrow B_3 B_4}+C^T_{B_1B_2\rightarrow B_3 B_4}\mbox{\boldmath $\sigma$}_1\cdot\mbox{\boldmath $\sigma$}_2
-N_{B_1B_3\phi}N_{B_2B_4\phi}
   \frac{(\mbox{\boldmath $\sigma$}_1\cdot\mbox{\boldmath $q$})(\mbox{\boldmath $\sigma$}_2\cdot\mbox{\boldmath $q$})}
        {\mbox{\boldmath $q$}^2+m^2-i\epsilon}\mathcal{I}_{B_1B_2\rightarrow B_3B_4}\, ,
\end{equation}
where $C^S$ and $C^T$ are linear combinations of $C_i^m$'s and $N_{BB'\phi}$ is determined by the initial-/final-state  baryons and the exchanged pseudoscalar meson~\cite{deSwart:1963pdg},
\begin{equation}
\begin{array}{lllll}
N_{NN\pi}  = f, & & N_{NN\eta}  = \frac{1}{\sqrt{3}}(4\alpha -1)f, & & N_{\Lambda NK} = -\frac{1}{\sqrt{3}}(1+2\alpha)f, \\
N_{\Lambda\Sigma\pi} = \frac{2}{\sqrt{3}}(1-\alpha)f, & & N_{\Lambda\Lambda\eta}  = -\frac{2}{\sqrt{3}}(1-\alpha )f, & & N_{\Sigma NK} = (1-2\alpha)f, \\
N_{\Sigma\Sigma\pi}  = 2\alpha f, & & N_{\Sigma\Sigma\eta}  = \frac{2}{\sqrt{3}}(1-\alpha )f,
\end{array}
\label{SU3f}
\end{equation}
where $\alpha = F/(F+D), f = g_A/(2f_0)$. The isospin factors $\mathcal{I}_{B_1B_2\rightarrow B_3B_4}$ are listed in Table \ref{IFOME}. $m$ is the mass of the exchanged pseudoscalar meson, and $\mbox{\boldmath $q$}$ is the momentum transfer in the center-of-mass frame.

\begin{table}[h]
\footnotesize
\renewcommand{\arraystretch}{1.1}
\centering
\caption{Isospin factors $\mathcal{I}_{B_1B_2\rightarrow B_3B_4}$ for the one-pseudoscalar-meson-exchange diagrams.
}
\label{IFOME}
\begin{tabular}{L{2.5cm}L{1.5cm}L{1.5cm}L{1.4cm}R{0.3cm}}
\hline\hline
 Channel     & Isospin & $\pi$ & $K$ & $\eta$ \\
\hline
 $\Lambda N \rightarrow \Lambda N$ & $\frac{1}{2}$ & $ 0$ & $1$ & $1$ \\
 $\Lambda N \rightarrow \Sigma N$  & $\frac{1}{2}$ & $-\sqrt{3}$ & $-\sqrt{3}$ & $0$ \\
 $\Sigma N \rightarrow \Sigma N$   & $\frac{1}{2}$ & $-2$ & $-1$ & $1$ \\
 $\Sigma N \rightarrow \Sigma N$   & $\frac{3}{2}$ & $1$ & $2$ & $1$   \\
\hline\hline
\end{tabular}
\end{table}

\begin{figure}
  \centering
  \includegraphics[width=0.15\textwidth]{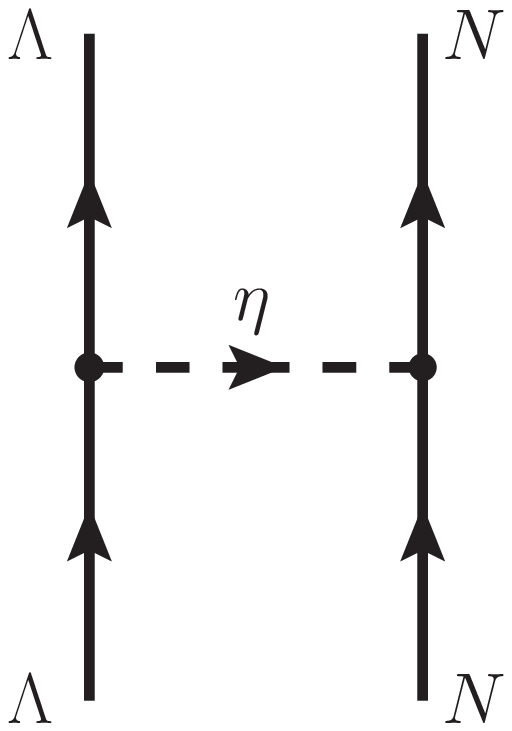}~~
  \includegraphics[width=0.15\textwidth]{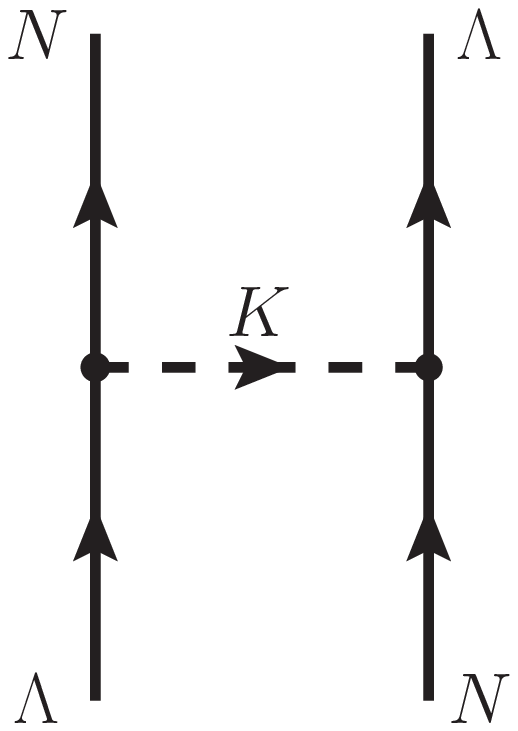}~~
  \includegraphics[width=0.15\textwidth]{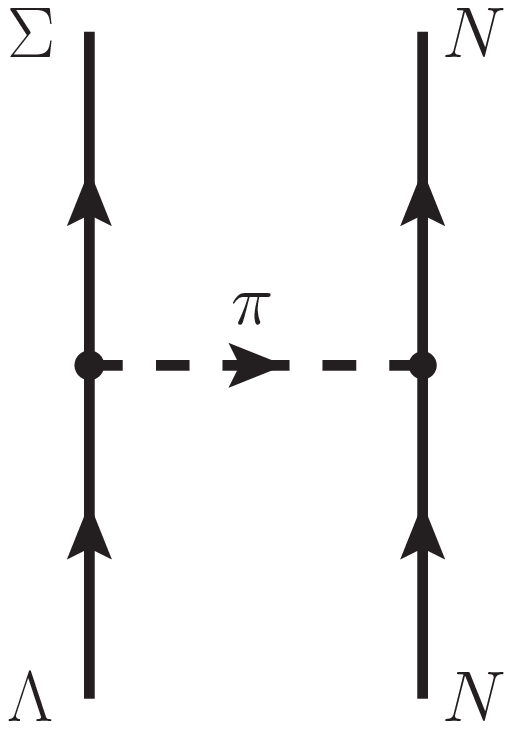}~~
  \includegraphics[width=0.15\textwidth]{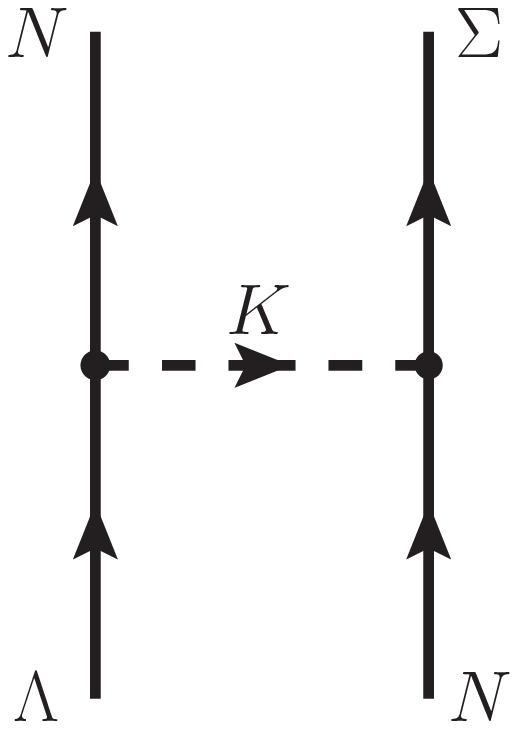}~~
  \includegraphics[width=0.15\textwidth]{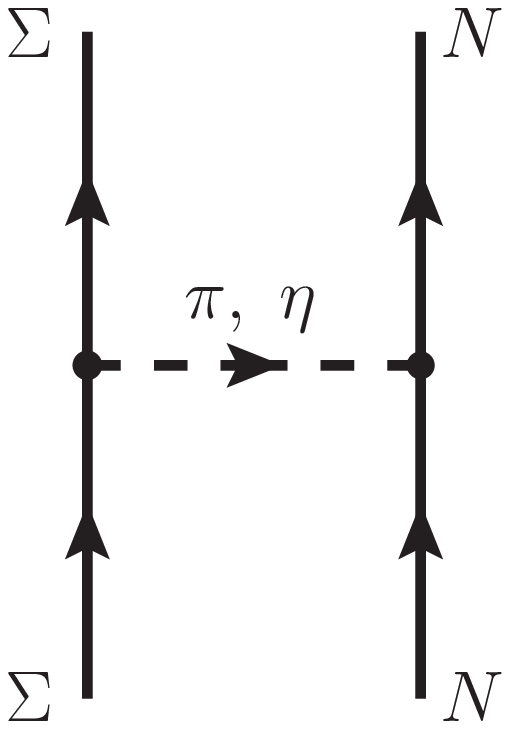}~~
  \includegraphics[width=0.15\textwidth]{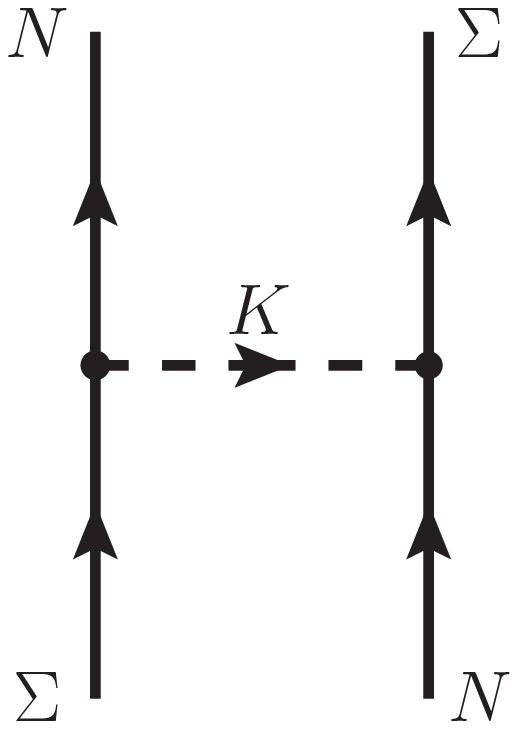}~~
  \caption{One-pseudoscalar-meson-exchange diagrams for the $\Lambda N-\Sigma N$ system.}\label{OME6}
\end{figure}

We focus here on the strangeness $S=-1$ $\Lambda N$-$\Sigma N$ sector. The relevant Feynman diagrams for LO potentials are depicted in Figs.~\ref{CT3} and~\ref{OME6}, in which the intermediate baryons and exchanged pseudoscalar mesons are identified.
After partial-wave projection, the contributions of contact terms can be expressed by five independent LECs, $C^{\Lambda \Lambda}_{1S0}$, $C^{\Sigma \Sigma}_{1S0}$, $C^{\Lambda \Lambda}_{3S1}$, $C^{\Sigma \Sigma}_{3S1}$, and $C^{\Lambda \Sigma}_{3S1}$, which need to be determined by fitting to the experimental data.

\subsection{Scattering equation}

The infrared enhancement in multibaryon propagations gives the theoretical argument for low-energy baryon-baryon interactions being nonperturbative~\cite{Weinberg:1991um}, in addition to the obvious phenomenological evidence that there exists a large number of mutlinucleon bound states---atomic nuclei. (The existence of exotic dibaryons, such as the $H$ dibaryon~\cite{Jaffe:1976yi} and the $N\Omega$ dibaryon~\cite{Goldman:1987ma}, has received much attention but has not been firmly confirmed.) As a result, one needs to iterate at least the LO ChEFT potentials. In the HB approach, the Lippmann-Schwinger equation with a nonrelativistic propagator is used,
 \begin{align}\label{SELS}
  T_{\rho\rho'}^{\nu\nu',J}(p,p';\sqrt{s})
  &=
  V_{\rho\rho'}^{\nu\nu',J}(p,p') \nonumber\\
  &\ \ \ \ +
  \sum_{\rho'',\nu''}\int_0^\infty \frac{dp''p''^2}{(2\pi)^3}
  V_{\rho\rho''}^{\nu\nu'',J}(p,p'')
  \frac{2\mu_{\nu''}}{q_{\nu''}^2-{p''}^2+i\epsilon}
  T_{\rho''\rho'}^{\nu''\nu',J}(p'',p';\sqrt{s})\, ,
\end{align}
where $\sqrt{s}$ is the total energy of the baryon-baryon system in the center-of-mass frame,  $q_{\nu''}$ is the relativistic on-shell momentum defined by $\sqrt{s} = \sqrt{M_{B_{1,\nu''}}^2+ q_{\nu''}^2} + \sqrt{M_{B_{2,\nu''}}^2+ q_{\nu''}^2}$, where $B_{1,\nu''}$ and $B_{2,\nu''}$ are intermediate state baryons, and $\mu_{\nu''}$ is the reduced mass  of  the intermediate state.
The labels $\nu,\nu',\nu''$ denote the particle channels, e.g., $\Lambda p$, $\Sigma^+ n$, $\Sigma^0 p $, and $\rho,\rho',\rho''$ denote the partial waves, e.g., $^1S_0$, $^3S_1$, etc.

The key to the EG approach is to use the Kadyshevsky equation to iterate the LO potentials so that as many relativistic effects as possible are included in the two-baryon propagation. The Kadyshevsky equation is one way to reduce the Bethe-Salpeter equation to a three-dimension form, first proposed in Ref.~\cite{Kadyshevsky:1967rs} (see Ref.~\cite{Woloshyn:1974wm} for other choices of reduction). In the context of $YN$ scattering, the equation can be written with notation similar to Eq.~\eqref{SELS}:
\begin{align}\label{SEK}
   T_{\rho\rho'}^{\nu\nu',J}(p,p';\sqrt{s})
  &=
   V_{\rho\rho'}^{\nu\nu',J}(p,p') \nonumber\\
  &\ \ \ \ +
  \sum_{\rho'',\nu''}\int_0^\infty \frac{dp''p''^2}{(2\pi)^3}
  \frac{2\mu_{\nu''}^2~ V_{\rho\rho''}^{\nu\nu'',J}(p,p'')~
  T_{\rho''\rho'}^{\nu''\nu',J}(p'',p';\sqrt{s})}{(p''^2+4\mu_{\nu''}^2)
  (\sqrt{q_{\nu''}^2+4\mu_{\nu''}^2}-\sqrt{p''^2+4\mu_{\nu''}^2}+i\epsilon)}.
\end{align}
It is a crucial difference that the propagator of the Kadyshevsky equation has in its denominator higher power of intermediate momentum $p''$ than that of the Lippmann-Schwinger equation \eqref{SELS}. Therefore, it has promising potential to mitigate the cutoff sensitivity.

To properly account for physical thresholds and the Coulomb force in charged channels, e.g., $\Sigma^-p\rightarrow\Sigma^-p$, we solve the scattering equations in the particle basis for
both the Lippmann-Schwinger and Kadyshevsky equations, while the kernel is evaluated in the isospin basis. Relativistic kinematics is used throughout to relate the laboratory momenta to the center-of-mass momenta. The Coulomb interaction for charged channels is treated with the Vincent-Phatak method~\cite{Vincent:1974zz, Holzenkamp:1989tq, Walzl:2000cx}.

Since the Lippmann-Schwinger and Kadyshevsky equations do nothing but resum a certain class of diagrams, they in principle need to be regularized in order for the integration to be well defined, as in many field-theoretical calculations. Nonperturbative calculations are not amenable to dimensional regularization, so we turn to cutoff regularization. In solving both integral equations, we multiply the potentials by the following Gaussian form factor in momentum space, as it was done in Refs.~\cite{Polinder:2006zh, Haidenbauer:2013oca},
\begin{equation}\label{EF}
  f_{\Lambda_F}(p,p') = \exp \left[-\left(\frac{p}{\Lambda_F}\right)^{2n}-\left(\frac{p'}{\Lambda_F}\right)^{2n}\right] \, ,
\end{equation}
where $n=2$.

\section{Fitting procedure\label{sec_fitting}}

At LO, there are five LECs in the strangeness $S=-1$ sector, which need to be determined by fitting to the experimental data. We use the same set of low-energy $YN$ scattering data as used in Refs.~\cite{Rijken:1998yy, Haidenbauer:2013oca}. It contains 36 data, of which 35 are total cross sections of $\Lambda N$ and $\Sigma N$ reactions~\cite{SechiZorn:1969hk,Alexander:1969cx,Engelmann:1966,Eisele:1971mk} with the laboratory momentum smaller than approximately $300$ MeV/$c$, $P_\textrm{lab} < 300$ MeV/$c$. These reactions include $\Lambda p \rightarrow \Lambda p$, $\Sigma^+ p \rightarrow \Sigma^+ p$, $\Sigma^- p \rightarrow \Sigma^- p$, $\Sigma^- p \to \Lambda n$, and $\Sigma^- p \to \Sigma^0 n$.  The last datum is the $\Sigma^- p$ inelastic capture ratio at rest \cite{Hepp:1968zza}.

It is customary to take as a further constraint the empirical value of the hypertriton $^3_\Lambda$H binding energy \cite{Juric:1973zq, Davis:1991zpu}, which has been known to be crucial in fixing the relative strength of the $^1S_0$ and the $^3S_1$$ -$ $^3D_1$ contributions to $\Lambda p$ scattering. Because of the exploratory nature of this work, however, we are content with using the value of the $S$-wave $\Lambda p$ scattering length from Ref.~\cite{Polinder:2006zh}, which was shown to reproduce a reasonable value for the hypertriton binding energy~\cite{Nogga:2013pwa}.

In the charged channels $\Sigma^+ p \rightarrow \Sigma^+ p$ and $\Sigma^- p \rightarrow \Sigma^- p$, the experimental values for total cross
section~\cite{Eisele:1971mk} were obtained by an incomplete angular coverage
\begin{equation}\label{ccts}
  \sigma = \frac{2}{\textrm{cos}~\theta_{\textrm{max}}-\textrm{cos}~\theta_{\textrm{min}}}
  \int^{\textrm{cos}~\theta_{\textrm{max}}}_{\textrm{cos}~\theta_{\textrm{min}}}
  \frac{\textrm{d}~\sigma(\theta)}{\textrm{d}~\textrm{cos}~\theta}~\textrm{d}~\textrm{cos}~\theta\, ,
\end{equation}
where $\theta$ is the angle between incoming and outgoing $\Sigma^\pm$ in the center-of-mass system. The Coulomb scattering amplitude goes to infinity at the forward angle. Following Refs.~\cite{Rijken:1998yy, Haidenbauer:2013oca},
we use $\textrm{cos}~\theta_{\textrm{min}}=-0.5$ and $\textrm{cos}~\theta_{\textrm{max}}=+0.5$ in our calculations for these two channels, in order to stay as close as possible to the experimental setup. Total cross sections for other channels are evaluated without any cutoff on $\theta$~\cite{Holzenkamp:1989tq}, but only partial waves with $J\leq 2$ are accounted for.

The inelastic capture ratio at rest is defined as~\cite{Swart:1962ny}
\begin{align}\label{icr1}
  r_R = \frac{1}{4}r_{S=0} + \frac{3}{4}r_{S=1}\, ,
\end{align}
with
\begin{align}\label{icr2}
  r_{S=0,1} = \frac{\sigma_{S=0,1}^{\Sigma^-p\rightarrow\Sigma^0n}}{\sigma_{S=0,1}^{\Sigma^-p\rightarrow\Sigma^0n}
  + \sigma_{S=0,1}^{\Sigma^-p\rightarrow\Lambda n}}\left|_{P_{\Sigma^-}=0}\right.,
\end{align}
where $\sigma$ is the cross section of the corresponding channel and $S=0,1$ denotes the spin singlet $^1 S_0$ and the triplet ${}^3S_1 - {}^3D_1$, respectively. Following the common practice \cite{Rijken:1998yy, Haidenbauer:2013oca}, we calculate the cross sections at a small nonzero momentum, i.e., $P_{\Sigma^-}=10$ MeV/$c$.

The fit is performed by minimizing the $\chi^2$, which is defined as
\begin{equation}
  \chi^2=\sum_i\frac{(D_i(\mathrm{Exp.})-D_i (\mathrm{Theo.}))^2}{\Delta_i^2}\, ,\label{eqn_chi2}
\end{equation}
where $i$ enumerates the input data; $D_i(\mathrm{Exp.})$ and $D_i(\mathrm{Theo.})$ denote, respectively, the experimental and theoretical values for certain observables; and $\Delta_i$ is the experimental uncertainty.

\section{Results and discussions\label{sec_results}}

Before presenting and discussing the results, we would like to review the primary goal: to compare the EG and HB approaches, in terms of their ability to describe the hyperon-nucleon scattering data and their sensitivity to a varying ultraviolet cutoff. By doing so, we hope to shed more light on the impact of relativistic effects encoded in the covariant formulation.

We first determine for a given value of $\Lambda_F$ the best-fit values of LECs by minimizing the $\chi^2$, as defined in Eq.~\eqref{eqn_chi2}, and then use this set of LECs to generate for this $\Lambda_F$ the phase shifts of various channels of hyperon-nucleon scattering.

\begin{figure}
  \centering
  \includegraphics[width=0.5\textwidth]{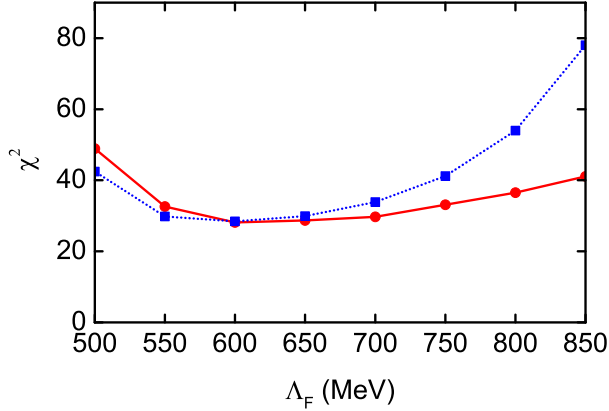}
  \caption{$\chi^2$ as a function of the cutoff in the EG approach (red solid line) and the HB approach (blue dotted line).}\label{Fig:chi2}
\end{figure}

In Fig.~\ref{Fig:chi2}, the $\chi^2$ is plotted as a function of $\Lambda_F$, in both EG and HB approaches, for the cutoff range $\Lambda_F = 500 - 850$ MeV. The optimum $\chi^2$ occur at similar cutoff values in two approaches, $\Lambda_F \simeq 600$ MeV, and their values appear to be almost identical. The values of the $\chi^2$ and LECs at $\Lambda_F = 600$ MeV are listed in Table~\ref{tab_LECs600}. We also observe that, as the cutoff increases, the quality of fit deteriorates much faster in the HB approach than in the EG approach. As we will see later, this rapid increase of the HB $\chi^2$ beyond $\Lambda \simeq 700$ MeV is intimately related to the limit-cycle-like cutoff dependence in attractive, triplet channels, such as $^3P_0$.

% \begin{figure}[htpb]

% \begin{table}[htpb]
\begin{table}
\centering
 \caption{Best-fitted values of $YN$ $S$-wave LECs (in units of $10^4$ GeV$^{-2}$) and $\chi^2$ for $\Lambda_F=600$ MeV in the EG and HB approaches.}
   \begin{tabular}{l|cccccc}
  \hline\hline
  &$\chi^2$ & $C^{\Lambda \Lambda}_{1S0}$ & $C^{\Sigma \Sigma}_{1S0}$ & $C^{\Lambda \Lambda}_{3S1}$ & $C^{\Sigma \Sigma}_{3S1}$ & $C^{\Lambda \Sigma}_{3S1}$ \\
  \hline
~~EG~~ & $28.23$ & $-0.04795(151)$ & $-0.07546(81)$ & $-0.01727(124)$ & $0.36367(30310)$ & $0.01271(471)$\\
~~HB~~ &$28.52$ & $-0.03894(1)$ & $-0.07657(1)$ & $-0.01629(13)$ & $0.20029(14050)$ &$-0.00176(304)$ \\
  \hline\hline
\end{tabular}\label{tab_LECs600}
 \end{table}

\begin{figure}[htpb]
  \centering
  \includegraphics[width=1.0\textwidth]{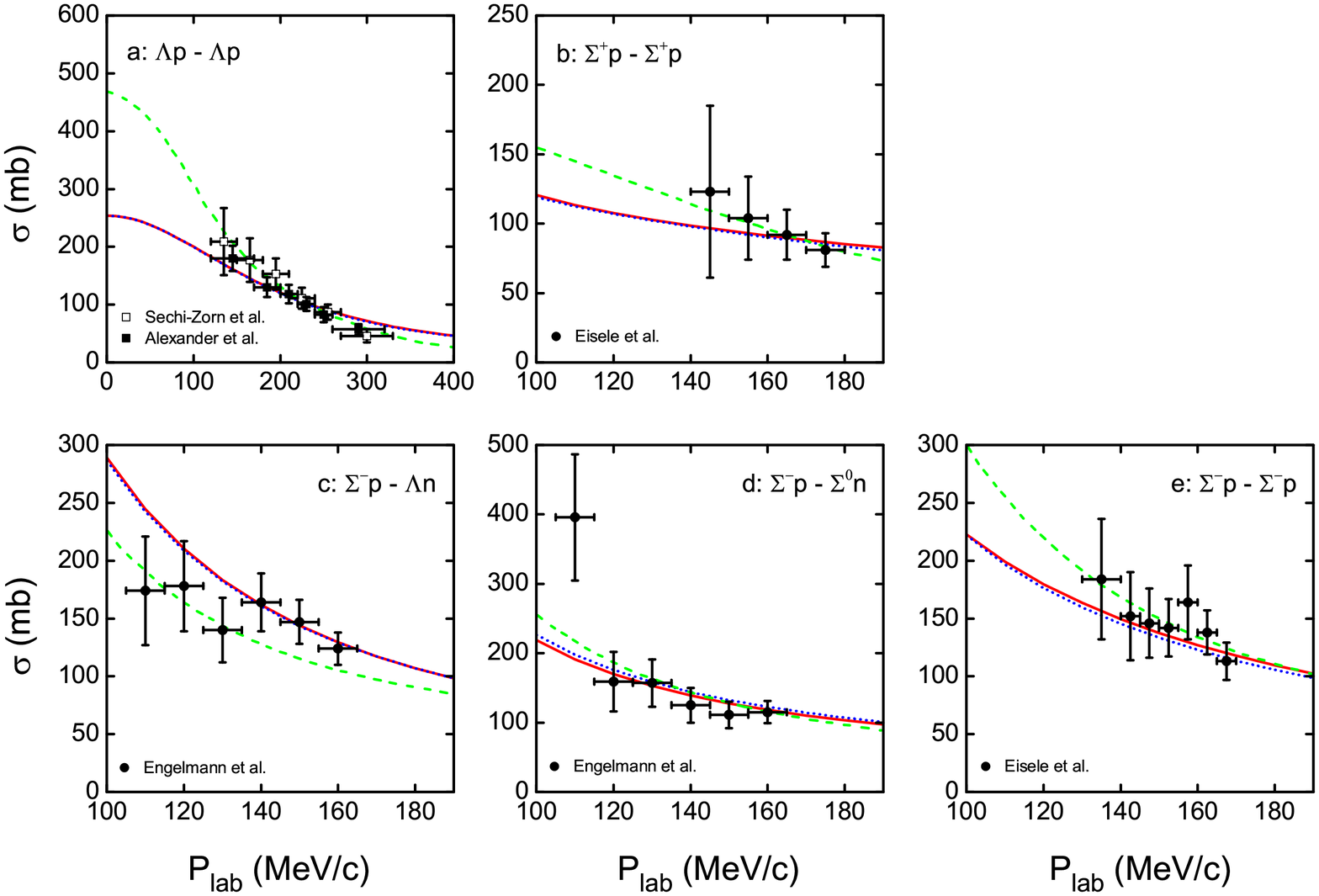}
  \caption{Cross sections in the EG approach (red solid lines) and the HB approach (blue dotted lines) as functions of the laboratory momentum, in comparison with the experimental data. For reference, the J\"ulich 04 results~\cite{Haidenbauer:2005zh} are also shown (green dashed lines). The experimental data are taken from Sechi-Zorn \textit{et al}.~\cite{SechiZorn:1969hk}, Alexander \textit{et al}.~\cite{Alexander:1969cx},
  Eisele \textit{et al}.~\cite{Eisele:1971mk},  and Engelmann \textit{et al}.~\cite{Engelmann:1966}.}\label{Fig:fitC}
\end{figure}

\begin{figure}[htpb]
  \centering
  \includegraphics[width=1.0\textwidth]{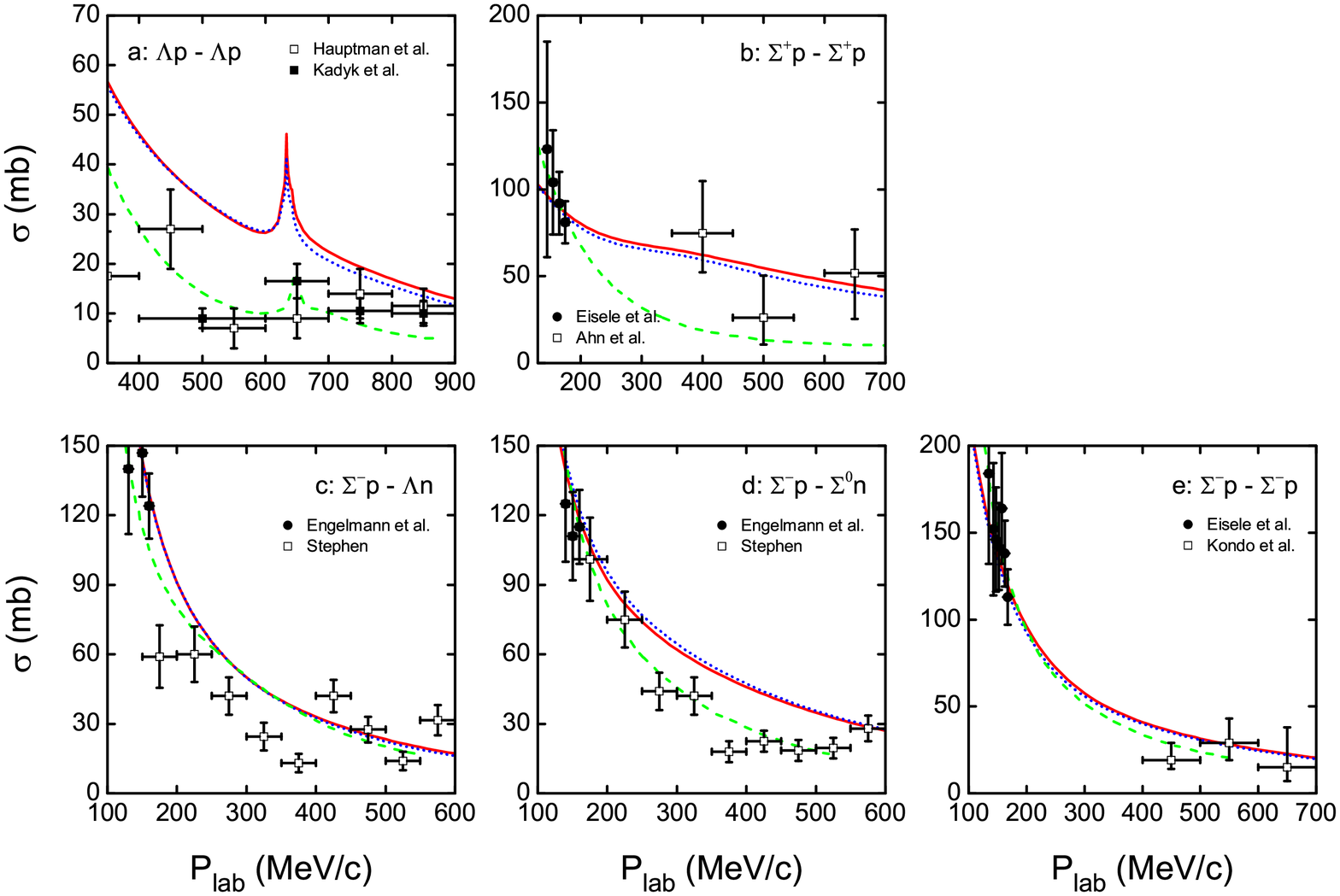}
  \caption{Predicted cross sections in the EG (red solid lines) and HB (blue dotted lines) approaches at $\Lambda_F = 600$ MeV, in comparison with the experimental values and the J\"ulich 04 results~\cite{Haidenbauer:2005zh} (green dashed lines). The experimental data
  are taken from Hauptman \textit{et al}.~\cite{Hauptman:1977hr}, Kadyk \textit{et al}.~\cite{Kadyk:1971tc}, Eisele \textit{et al}.~\cite{Eisele:1971mk}, Ann \textit{et al}.~\cite{Ahn:2005gb}, Engelmann \textit{et al}.~\cite{Engelmann:1966}, Stephen~\cite{stephen}, and Kondo \textit{et al}.~\cite{Kondo:2000hn}.}\label{Fig:preC}
\end{figure}

In Fig.~\ref{Fig:fitC}, we compare the cross sections as functions of $P_\textrm{lab}$, calculated with LECs shown in Table~\ref{tab_LECs600}, to low-energy experimental values that are included in the fitting. Consistent with the $\chi^2$ plot in Fig.~\ref{Fig:chi2}, the two approaches yield basically the same results. We also make predictions for higher $P_\textrm{lab}$, as shown in Fig.~\ref{Fig:preC}, and once again the curves for both approaches are identical, in comparison with the experimental uncertainties.

A short summary is in order before we proceed. Figures.~\ref{Fig:chi2}, \ref{Fig:fitC}, and \ref{Fig:preC} tell us that if we are given the freedom to choose an optimum value for $\Lambda_F$, we will find that the EG and HB approaches describe the data with similar quality, although the EG approach is less sensitive to the cutoff when $\Lambda_F > 600$ MeV.

At the LO of nucleon-nucleon scattering, it has been shown that the cutoff sensitivity is mostly caused by the singular attraction of one-pion exchange (OPE)~\cite{PavonValderrama:2005gu, Birse:2005um, Nogga:2005hy}. Since whether OPE is attractive or repulsive depends on the matrix element of the tensor projector between partial waves, it is necessary to investigate the cutoff dependence of individual partial-wave amplitudes. In hyperon-nucleon scattering, we will also look into the cutoff dependence of partial-wave phase shifts, for OPE has a similar structure to OPME. For definitiveness, $\Lambda p$ and $\Sigma^+ p$  scatterings will be considered.

With LECs determined by the aforementioned fitting procedure, the $S$-, $P$-, and $D$-wave phase shifts and the mixing angles for ${}^3S_1 - {}^3D_1$ ($\epsilon_1$) and ${}^3P_2 - {}^3F_2$ ($\epsilon_2$) are calculated as functions of the cutoff, for two values of $P_\textrm{lab}$: $P_\textrm{lab} = 300$ and $900$ MeV/$c$. At $P_\textrm{lab} = 300$ MeV/$c$, only the $\Lambda p$ channel is  physically open, while at $P_\textrm{lab} = 900$ MeV/$c$, all three coupled channels ($\Lambda p$, $\Sigma^+ n$, $\Sigma^0 p$) are open.  Note that $\Lambda_F$ constrains more directly the values of the center-of-mass momentum, so $P_\textrm{lab} = 900$ MeV/$c$ is not necessarily a concern for the cutoff values smaller than $900$ MeV. For $\Lambda p$ scattering, $P_\textrm{lab} = 900$ MeV/$c$ corresponds to the center-of-mass momentum of the proton being $P_\textrm{CM} = 384.8$ MeV/$c$.

\begin{figure}[t]
  \centering
  \includegraphics[width=1.0\textwidth]{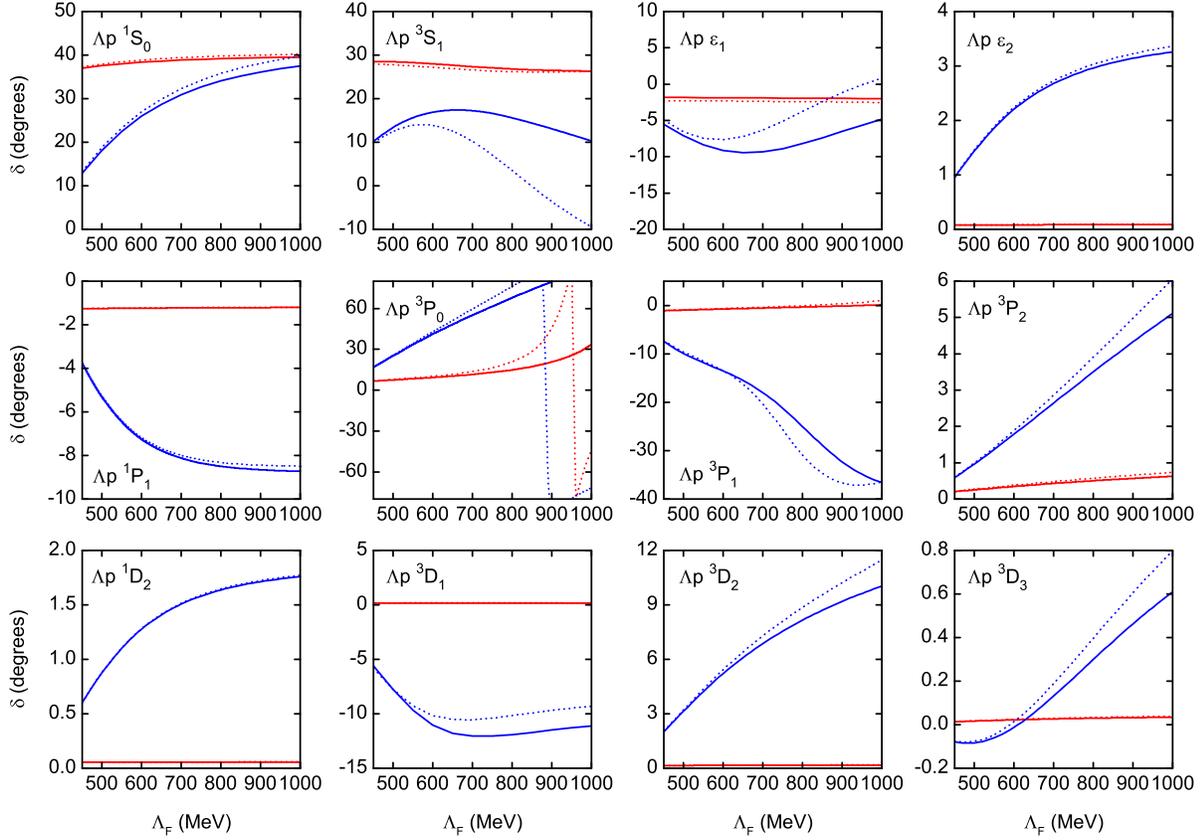}
  \caption{$S$, $P$, and $D$-wave phase shifts and ${}^3S_1 - {}^3D_1$ (~${}^3P_2 - {}^3F_2$) mixing angles $\epsilon_1$ ($\epsilon_2$) for $\Lambda p$ scattering, as a function of the cutoff $\Lambda_F$. The solid (dotted) lines correspond to the EG (HB) approach. Red (blue) lines represent $P_\textrm{lab} = 300$ ($900$) MeV/$c$.}\label{Lap_PDFL}
\end{figure}

\begin{figure}[t]
  \centering
  \includegraphics[width=1.0\textwidth]{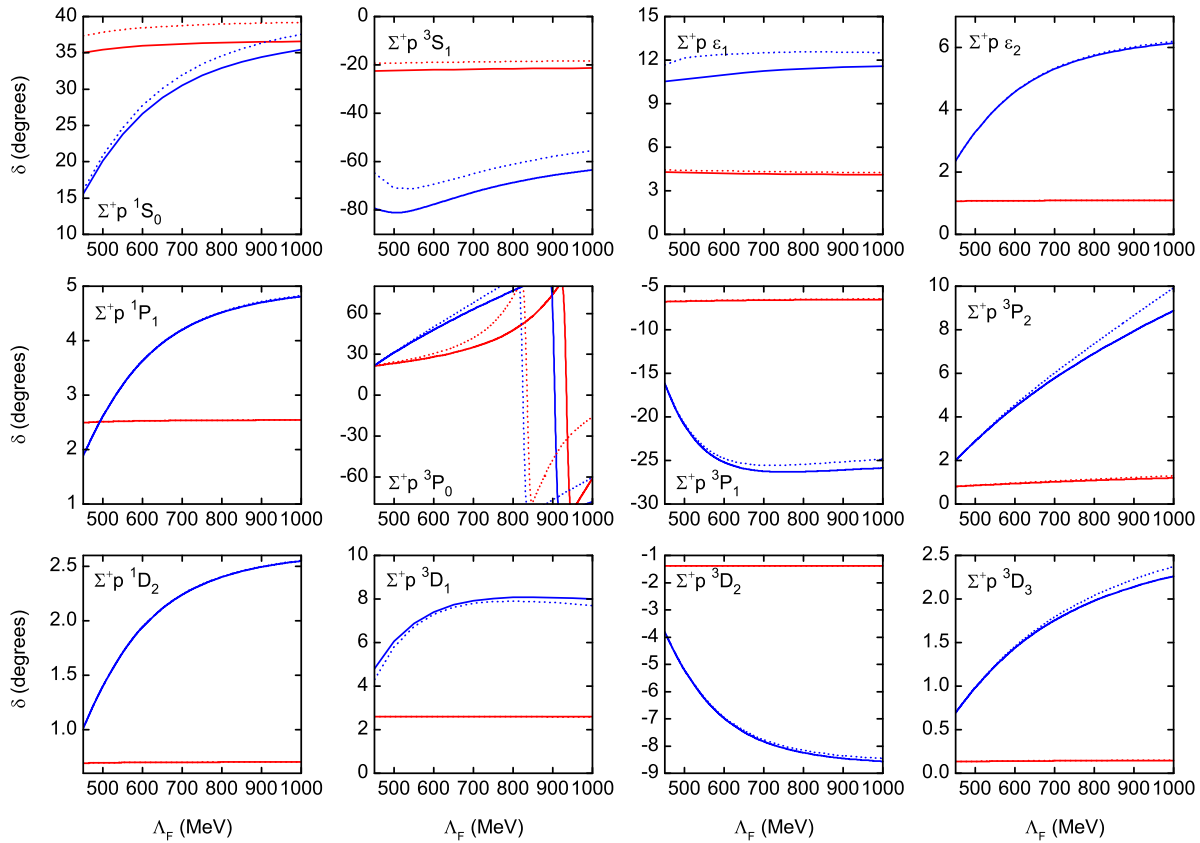}
  \caption{$S$, $P$, and $D$-wave phase shifts and ${}^3S_1 - {}^3D_1$ (~${}^3P_2 - {}^3F_2$) mixing angles $\epsilon_1$ ($\epsilon_2$) for $\Sigma^+ p$ scattering, as a function of the cutoff $\Lambda_F$.
  % $S$-, $P$- and $D$-wave phase shifts and mixing angles as a function of the cutoff in the region of 450-1000 MeV for the $\Sigma^+ p$ channel.
 Symbols are the same as those of Fig.~\ref{Lap_PDFL}.}\label{Sip_PDFL}
\end{figure}

\begin{figure}[t]
  \centering
  \includegraphics[width=1.0\textwidth]{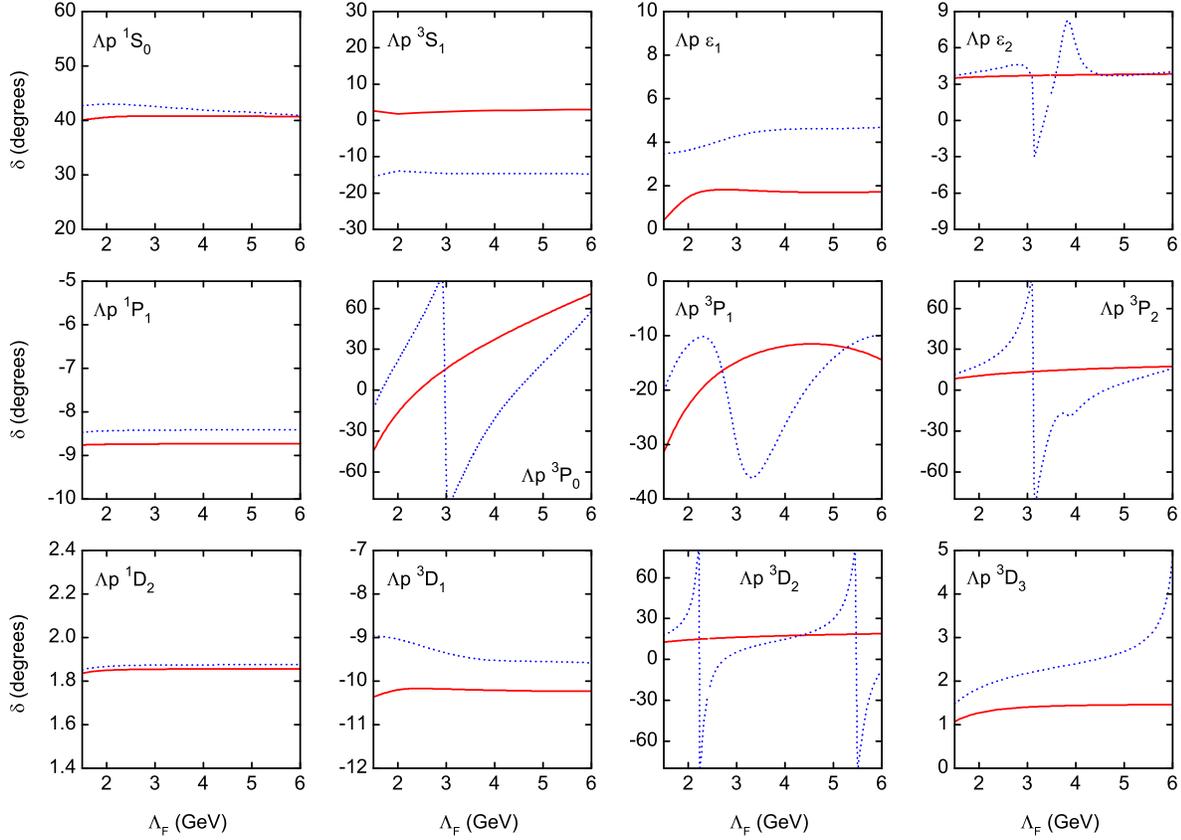}
  \caption{Phase shifts and mixing angles of $\Lambda p$ scattering as a function of the cutoff in the region $\Lambda_F = 1.5 - 6$ GeV. The solid (dotted) lines correspond to the EG (HB) approach.\label{Lap_PDFH}}
  % The solid and dotted lines denote the results obtained by the EG approach and the Weinberg approach at $P_\textrm{lab} = 900$ MeV.}
\end{figure}

\begin{figure}[t]
  \centering
  \includegraphics[width=1.0\textwidth]{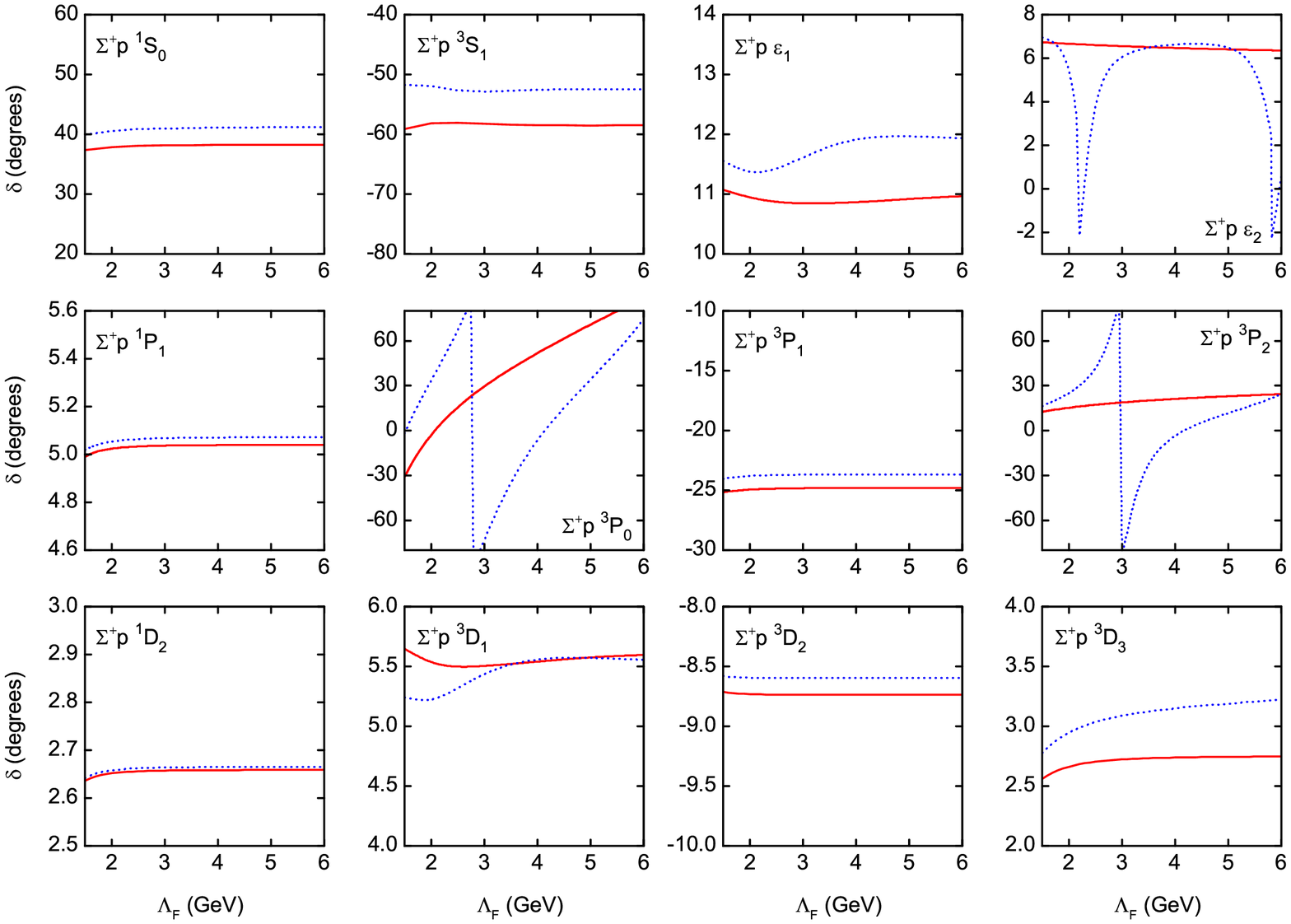}
  \caption{Phase shifts and mixing angles of $\Sigma^+ p$ scattering as a function of the cutoff in the region $\Lambda_F = 1.5 - 6$ GeV. Symbols are the same as those of Fig.~\ref{Lap_PDFH}.}\label{Sip_PDFH}
\end{figure}

We split the cutoff range into two parts to study the cutoff sensitivity. One is $\Lambda_F = 450 - 1000$ MeV, which is more conventionally used for practical calculations. The other is $\Lambda_F = 1.5 - 6$ GeV, where the cutoff dependence is more explicitly probed. Results with softer cutoffs are shown in Figs.~\ref{Lap_PDFL} and \ref{Sip_PDFL}, and harder cutoffs are used in Figs.~\ref{Lap_PDFH} and ~\ref{Sip_PDFH}.

\begin{figure}[htpb]
  \centering
  \includegraphics[width=0.45\textwidth]{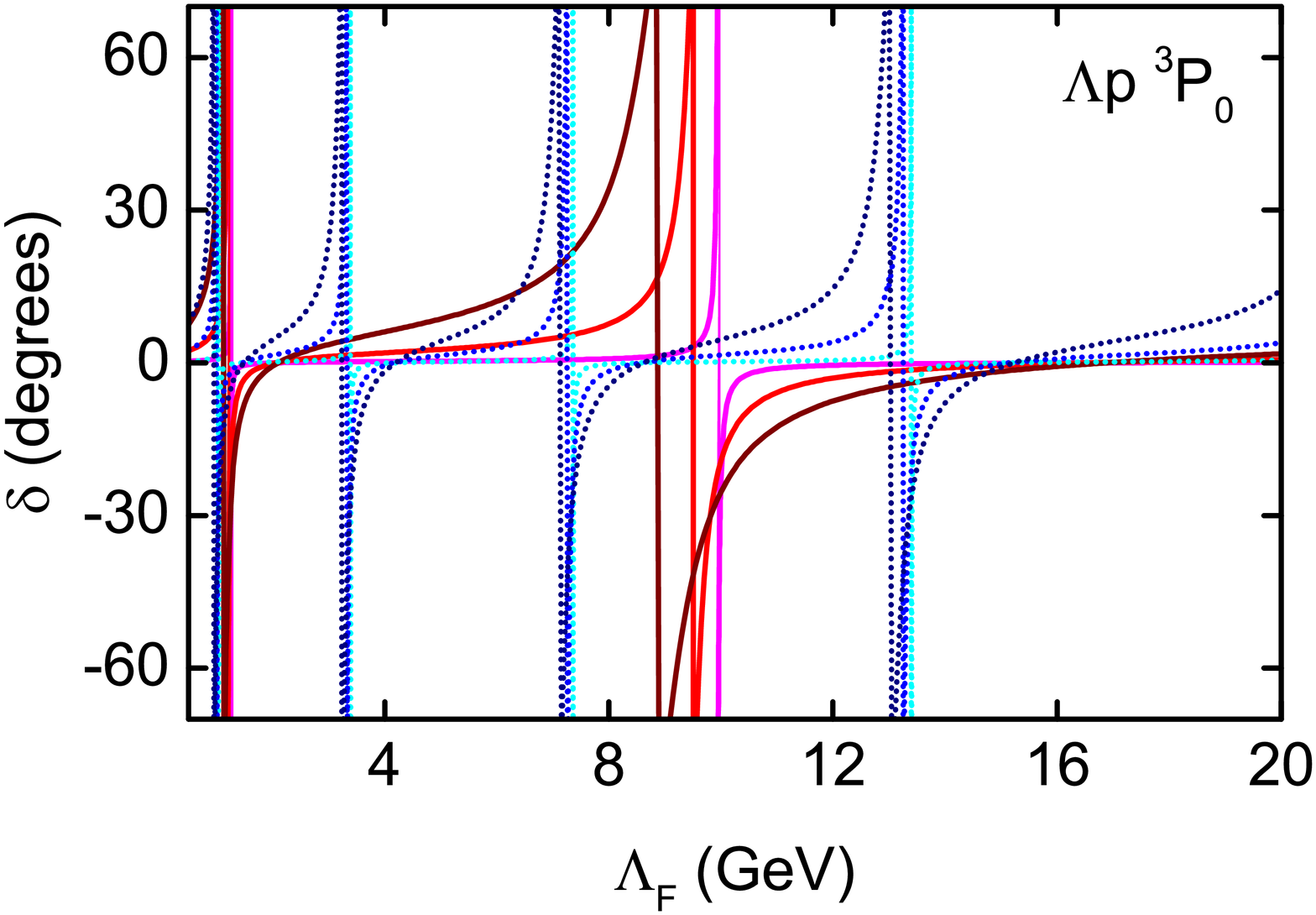}
    \includegraphics[width=0.45\textwidth]{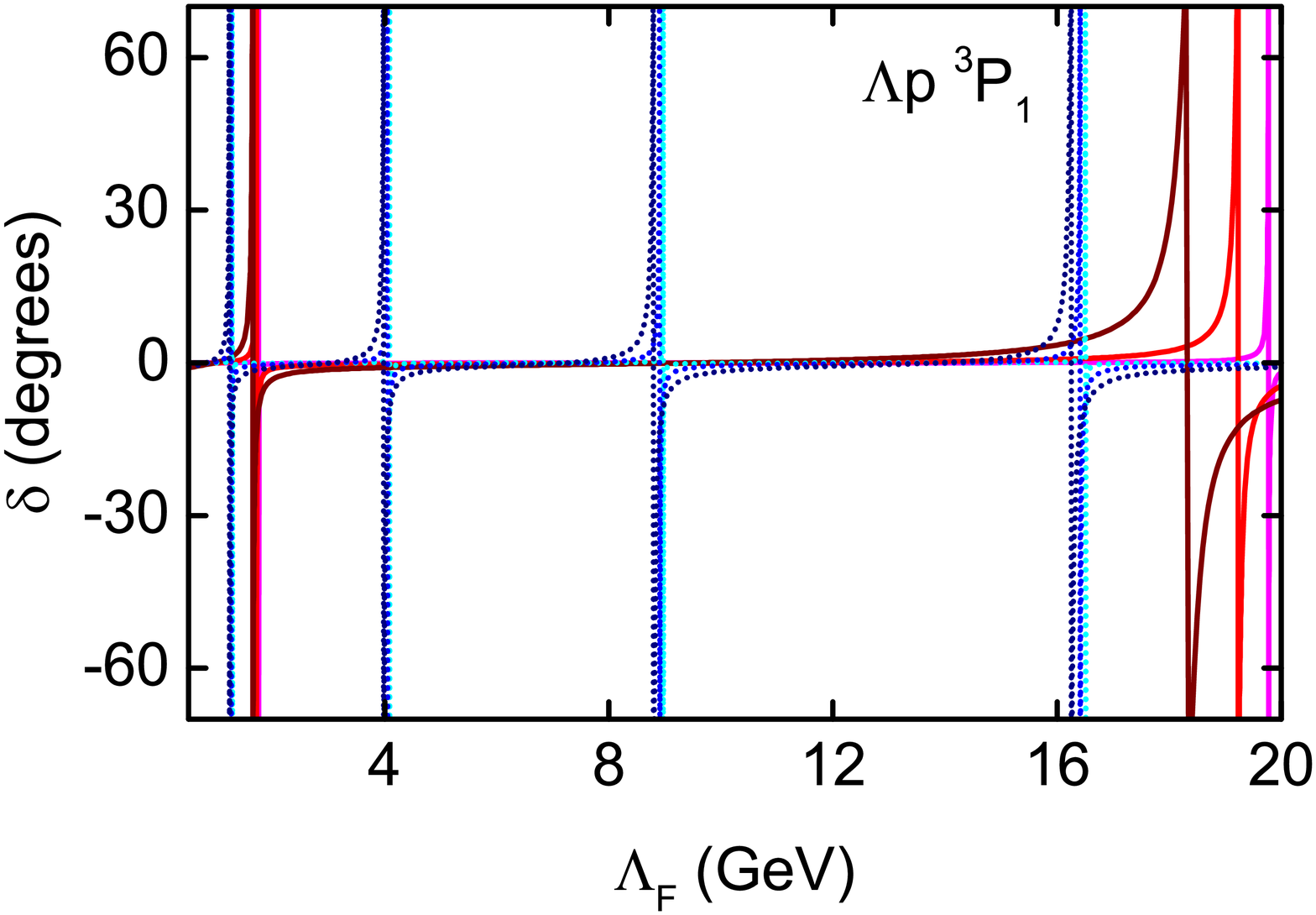}
  \caption{$^3P_0$ and $^3P_1$ phase shifts of $\Lambda p$ scattering at $P_\mathrm{lab}=100$ (magenta and cyan), 200 (red and blue), 300 (wine and navy) MeV/$c$ as a function of the cutoff, calculated in the EG (solid lines) and HB (dotted lines) approaches.}\label{Fig:chi2Le}
\end{figure}

Similar to $NN$ scattering~\cite{Epelbaum:2012ua}, the EG approach removes limit-cycle-like cutoff dependence from some of the partial waves. Let us first look at $\Lambda p$ scattering. With the HB approach, significant cutoff variations are present in phase shifts of ${}^3P_0$, ${}^3P_1$, ${}^3P_2$, ${}^3D_2$, $^3D_3$, and the mixing angle $\epsilon_2$. Switching to the EG approach suppresses greatly phase-shift oscillations in ${}^3P_2$, ${}^3D_2$, $^3D_3$, and $\epsilon_2$. Although ${}^3P_0$ and ${}^3P_1$ remain sensitive to cutoff variation, the cutoff period of cycles becomes generally wider, consistent with the general expectation that the cutoff dependence is mitigated when the EG approach is used. (To be certain about ${}^3P_1$, we show in Fig.~\ref{Fig:chi2Le} the cutoff dependence of its phase shifts at various $P_\textrm{lab}$ up to $20$ GeV.)

The similar pattern applies to $\Sigma^+ p$ scattering, with, however, an important difference. There are fewer problematic partial waves even in the HB approach. We see limit-cycle-like cutoff dependence in only ${}^3P_0$, ${}^3P_2$, and the mixing angle $\epsilon_2$, and the EG approach removes the cutoff sensitivity in ${}^3P_2$ and $\epsilon_2$.

\begin{figure}[htpb]
  \centering
  \includegraphics[width=0.5\textwidth]{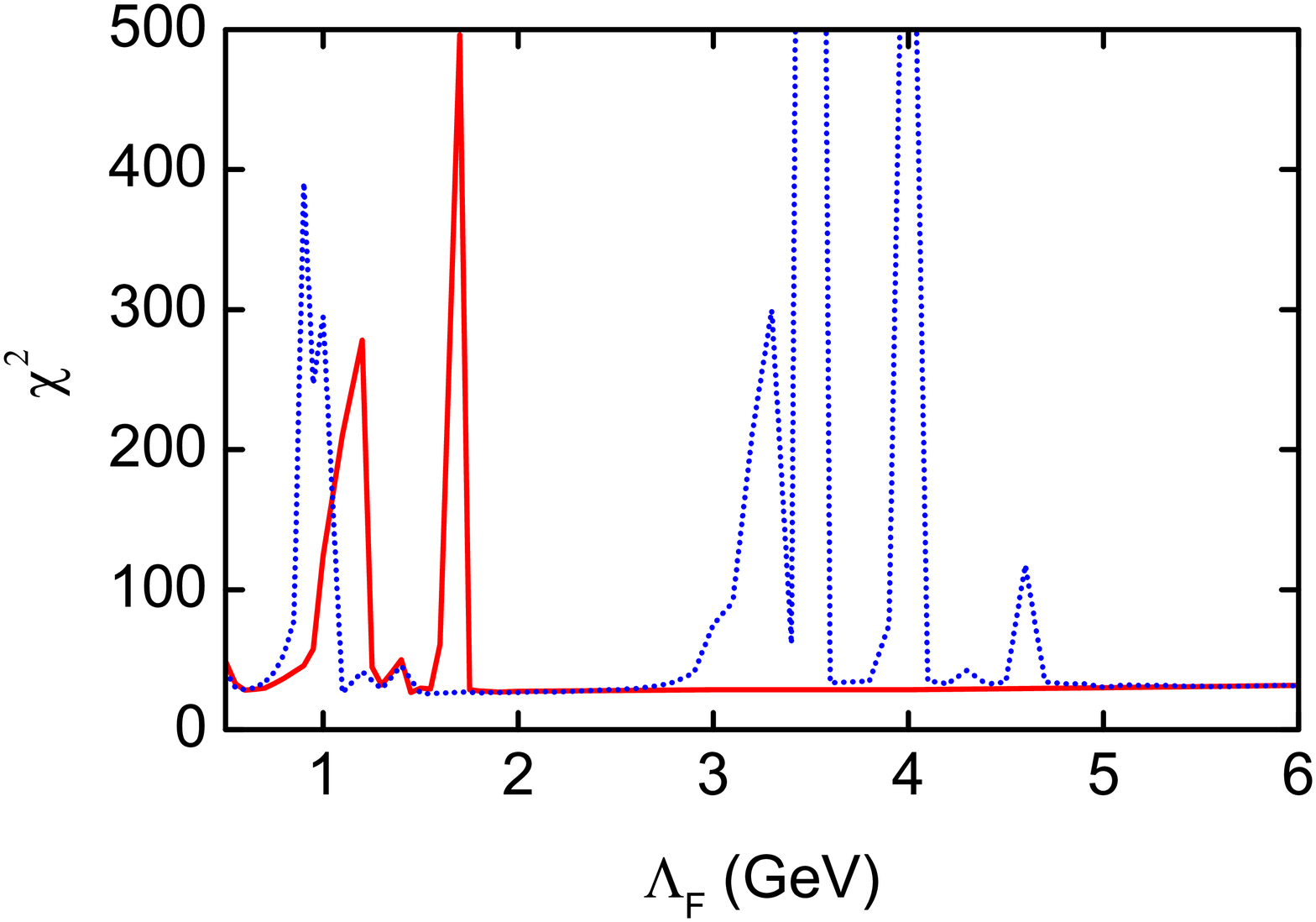}
  \caption{The best-fitted $\chi^2$ as a function of the cutoff in the region $\Lambda_F = 0.5 - 6$ GeV in the EG (red solid line) and HB (blue dotted line) approaches.}\label{Fig:chi2L}
\end{figure}

The limit-cycle-like cutoff dependence at hard cutoff values in certain partial waves prompts us to investigate how the best-fitted $\chi^2$ behaves at those cutoff values. In Fig.~\ref{Fig:chi2L}, the $\chi^2$ is plotted as a function of $\Lambda_F$ up to 6 GeV. After several peaks between $1$ and $2$ GeV, the EG $\chi^2$ remains almost constant with a value close to that obtained at $\Lambda_F = 600$ MeV. This is not the case, however, with the HB approach, of which the $\chi^2$ has a few more spikes from $3$ to $5$ GeV.

This distinctive difference in the $\chi^2$ between the two approaches must originate from the limit-cycle-like behavior in the phase shifts. We choose $^3P_0$ and $^3P_1$ of $\Lambda p$ scattering, shown in Fig.~\ref{Fig:chi2Le}, to reflect the correlation between the $\chi^2$ and the phase shifts. The $^3P_0$ and the $^3P_1$ partial-wave phase shifts are rather small at relevant $P_\mathrm{lab}$ for most cutoff values; therefore, they contribute little to the cross section. However, their contributions take a hike when the phase shifts cross the transition point of limit cycles with respect to the cutoff. Such a behavior exists in both
EG and HB approaches, except that the EG cycles are much wider. For particular values of $P_\mathrm{lab} = 100, 200, 300$ MeV/$c$ (close to energies of data points), only one cycle is observed with the EG approach in the cutoff range of $0.5 - 6$ GeV, but about 2.5 cycles are seen with the HB approach. This partially explains the peculiar dependence of the $\chi^2$ on the cutoff shown in Fig.~\ref{Fig:chi2L}.

\section{Summary and outlook\label{sec_summary}}

We have studied hyperon-nucleon scattering with strangeness $S=-1$ in covariant chiral effective field theory at LO, assuming NDA for the power counting of baryon-baryon contact operators. The focus has been on the comparison between the more conventional, heavy-baryon approach and the covariant-baryon approach proposed by Epelbaum and Gegelia, in terms of their cutoff sensitivity and their ability to describe the hyperon-nucleon data.

For each cutoff we looked at, we first determined the values of LECs---couplings of five contact operators---by minimizing the $\chi^2$ of 36 data, and then generated the phase shifts of $\Lambda p$ and $\Sigma^+ p$ scattering. The first finding is that if we are allowed to choose an optimum value of $\Lambda_F$ in fitting, there is not much difference between the two approaches, as far as the fit quality goes; the $\chi^2$ converges to almost an identical value at $\Lambda \simeq 600$ MeV.

The phase shifts were then investigated to expose the origin of cutoff dependence. In general, the EG approach mitigates the cutoff dependence, in comparison with the HB approach. More specifically, it removes from ${}^3P_2$, ${}^3D_2$, ${}^3D_3$, the mixing angle $\epsilon_2$ of $\Lambda p$ scattering, and ${}^3P_2$ and $\epsilon_2$ of $\Sigma^+ p$ scattering the limit-cycle cutoff dependence that existed in the HB approach. However, a significant cutoff sensitivity still persists in ${}^3P_0$ and ${}^3P_1$ of $\Lambda p$ and ${}^3P_0$ of $\Sigma^+ p$.

$SU(3)$ flavor symmetry was enforced upon the LO contact operators---all responsible for $S$ waves---but the long range potential OPME has $SU(3)$ breaking effects incorporated, e.g., the mass splitting of the Goldstone mesons. Since we did not see from the numerical results any $S$-wave cutoff dependence, it suggests at least in $S$ waves that $SU(3)$-violating counterterms are not needed for renormalization purposes. That is, the ${}^1S_0$ and ${}^3S_1$ counterterms of $\Lambda p$ and $\Sigma^+ p$ are correlated by $SU(3)$ symmetry, and even though the long-range potentials break $SU(3)$ symmetry, the counterterms appear to renormalize the amplitudes up to the cutoff values considered here. It is not clear to us that this observation will hold true if in a future work we promote ${}^3P_0$ and ${}^3P_1$ counterterms to LO to remove the cutoff dependence, along the line of thinking of Refs.~\cite{Nogga:2005hy, Birse:2005um}. The $SU(3)$ aspect of short-range interactions will be interesting to investigate, in light of renormalization.

\section{Acknowledgements}
 K. W. L. and L. S. G. thank Emiko Hiyama for her hospitality during their stay at RIKEN, where part of this work is done. K. W. L. acknowledges enlightening communications and discussions with
 Johann Haidenbauer, Stefan Petschauer, Jambul Gegelia, and Xian-Wei Kang.
This work is partly supported by the National Natural Science Foundation of China under Grants No. 11375024, No. 11522539, and No. 11375120 and the Fundamental Research Funds for the Central Universities.

\end{document}